\documentclass[12pt,draftcls,onecolumn]{IEEEtran}

\usepackage[T1]{fontenc}
\usepackage[utf8]{inputenc}
\usepackage{cite}
\usepackage{amsmath,amssymb}
\usepackage{graphicx}
\usepackage{booktabs}
\usepackage{tabularx}
\usepackage{array}
\usepackage{xcolor}
\usepackage{tikz}
\usepackage[most]{tcolorbox}
\usepackage{url}
\usepackage[hidelinks]{hyperref}
\usetikzlibrary{arrows.meta,positioning,shapes.geometric,calc}

\newcolumntype{Y}{>{\raggedright\arraybackslash}X}

\title{Ten Years of Deep Learning for Wireless Communications: From Learned Blocks to Deployable Wireless Intelligence}

\author{Hao~Ye,  Geoffrey~Ye~Li, and Biing-Hwang~Juang%
\thanks{H. Ye is with the Department of Electrical and Computer Engineering, Baskin School of Engineering, University of California, Santa Cruz, CA 95064 USA (e-mail: yehao@ucsc.edu).}
\thanks{G. Y. Li is with the Department of Electrical and Electronic Engineering, Imperial College London, London SW7 2AZ, U.K. (e-mail: geoffrey.li@imperial.ac.uk).}
\thanks{B.-H. Juang is with the School of Electrical and Computer Engineering, Georgia Institute of Technology, Atlanta, GA 30332 USA (e-mail: juang@ece.gatech.edu).}
}

\begin{document}
\maketitle

\begin{abstract}
Over the past decade, deep learning has evolved from a tool for replacing isolated wireless blocks into a broader methodology for developing wireless intelligence. This article traces that trajectory through three shifts: learning wireless functional modules, redesigning and re-normalizing communication goals, and enabling generalization under practical physical constraints. Together, these shifts advance the broader pursuit of communication anytime and anywhere, through any appropriate means. Early studies showed that neural networks could approximate difficult physical-layer inference and network-optimization mappings, while subsequent research embedded domain-specific structure, shifted toward task-oriented semantics, and addressed the need for edge-efficient adaptation. Looking ahead, we argue that the next era of wireless artificial intelligence (AI) depends on more than scaling model capacity. Promising directions include physically grounded wireless world models, agentic reasoning and fulfillment, and standardization mechanisms that allow learned components to operate with clear boundaries, physical consistency, and system-level interoperability.

\end{abstract}


\section{Introduction}

Wireless communication systems have historically relied on rigorous mathematical modeling, statistical signal processing, and information theory. This classical paradigm achieved remarkable success by decomposing complex architectures into well-defined functional blocks with explicit roles and interfaces. As networks grow in scale, density, and service diversity, however, many design problems have become difficult to capture with fixed analytical models or handcrafted algorithms alone. Model mismatch, hardware imperfections, dynamic interference, and non-stationary traffic all create regimes in which data-driven methods can complement traditional communication theory. Deep learning has therefore emerged as a flexible tool for learning complex relationships that are difficult to specify analytically.

Initial efforts in wireless deep learning centered on mapping approximation, using neural networks as efficient substitutes for selected inference and optimization procedures. Early studies showed that deep architectures could learn complex nonlinear mappings for physical-layer processing and wireless resource allocation~\cite{ye2018power,wen2018deep,sun2018learning}. These works established that trained models can absorb part of the complexity of difficult wireless functions during offline training and then provide fast online inference, making learning attractive for latency-constrained operation.

Subsequent research expanded the role of learning beyond approximating individual wireless functions, exploring how coupled communication components can be optimized jointly and how wireless design can be aligned with higher-level system objectives. This evolution led to end-to-end transceiver optimization, wherein the communication link is trained jointly as an autoencoder~\cite{oshea2017introduction,ye2020deep}. It also broadened the communication objective beyond intermediate block-level criteria: deep joint source--channel coding optimized source reconstruction over the complete link~\cite{bourtsoulatze2019deep}, semantic communication targeted task-relevant meaning~\cite{xie2021deep}, and wireless edge learning coupled distributed optimization with radio resource allocation~\cite{dinh2021federated}.

Despite these architectural and objective-driven advances, important limitations remain. Purely data-driven, task-specific models are often tied to the environments in which they are trained and may degrade under distribution shifts. Practical wireless systems also impose constraints that are rarely central in standard machine-learning benchmarks, including signaling overhead, tight decision latency, and limited energy. Frequent retraining or heavy model exchange can therefore reduce the net benefit of learning, leaving a gap between controlled simulation results and deployable wireless intelligence.

Looking ahead, wireless research is increasingly influenced by broader advances in artificial intelligence (AI), including foundation models, generative AI, and agentic systems. These paradigms suggest a path from isolated task predictors toward wireless models with stronger transfer, reasoning, and context awareness~\cite{guler2026foundation,liang2026llm}. The broader aspiration is communication anytime and anywhere, through any appropriate means. Agentic reasoning and fulfillment may extend intelligence across the communication process. At the same time, they must be reconciled with radio time scales, physical consistency, and standardized network operation. The next stage of wireless AI will therefore require not only more capable models, but also clearer principles for grounding, constraining, and integrating them within communication systems.

This article organizes the first decade of deep learning for wireless communications around three expanding roles: learning selected wireless functions, redesigning and re-normalizing communication goals, and enabling adaptive wireless intelligence under practical physical constraints. This progression is closely related to broader visions of AI-native air interfaces~\cite{hoydis2021toward}. Section~II reviews learned wireless modules, Section~III discusses goal-oriented communication, and Section~IV focuses on generalization, adaptation, and efficient edge execution. Section~V then identifies future research directions in standardization, multimodal wireless world models, and agentic reasoning and fulfillment. The central message is that the long-term impact of wireless AI will depend not on model capacity alone, but on whether learned models have clear roles, structural assumptions, interoperable interfaces, and well-understood operational limits within practical communication systems.

\section{Learning Wireless Functional Modules}

Deep learning was first adopted in wireless communications as a trainable implementation mechanism for selected functional modules. Classical transceiver and network design partitions a communication link into blocks, each governed by mathematical roles, algorithmic procedures, and standardized protocol interfaces. While this modularity remains indispensable for tractability and interoperability, it becomes restrictive when mathematical models are mismatched, inference rules are difficult to derive, or iterative optimizations are too slow for real-time execution. Deep learning introduced a complementary implementation principle: selected wireless functions can be represented by trainable mappings while preserving their roles within the communication architecture. By exploiting data rather than relying solely on fixed analytical models, these trainable mappings can better accommodate conditions that vary over time and across deployment environments, supporting communication anytime and anywhere.

\subsection{Physical-Layer Inference and Representation}

Physical-layer processing naturally involves inference and representation. Receivers infer transmitted information or channel states from observations shaped by propagation and hardware effects. Classical estimators and detectors exploit explicit statistical assumptions. When these assumptions hold, they provide strong performance and interpretable structure. However, when the assumptions do not hold, fixed algorithms may struggle to capture unmodeled impairments or empirical correlations.

Deep neural networks address this limitation by learning such mappings directly from examples. In orthogonal frequency-division multiplexing (OFDM) systems, neural networks can jointly support channel estimation and signal detection from received pilots and data symbols, showing robustness to imperfect channel knowledge and hardware nonlinearities~\cite{ye2018power}. The benefit is not only accuracy under a prescribed model: learning can absorb unmodeled physical effects, exploit empirical regularities, and combine multiple receiver operations into a compact offline-trained inference procedure. Learned channel state information (CSI) feedback offers a complementary perspective. In frequency-division duplex massive multiple-input multiple-output (MIMO) systems, CsiNet and related neural encoder--decoder models formulate channel state information feedback as a learned compression problem: the user equipment maps channel information to a compact latent code, and the base station reconstructs the representation needed for subsequent transmission decisions~\cite{wen2018deep}. This shifts feedback from a fixed compression-and-reconstruction pipeline to a trainable information bottleneck that preserves task-relevant channel information. These examples show that learning physical-layer modules is more than replacing one algorithm with another. The learned output must remain compatible with the surrounding communication procedures and preserve the information needed for subsequent transmission decisions.

\subsection{Resource Allocation and Network Decisions}

Deep learning also enabled network decision modules to be learned through amortized optimization. Wireless resource allocation is traditionally formulated through optimization problems whose objectives encode communication requirements. These formulations define network goals precisely, but the resulting problems are often nonconvex, high-dimensional, and coupled across network entities and time scales.

Deep learning reduces this burden by shifting repeated computation from online solving to offline training. Instead of running an expensive solver for every network state, a neural network can be trained to approximate the corresponding solution map. Learning-based power control exemplifies this paradigm, where a neural network emulates the decisions of the iterative weighted minimum mean-squared error (WMMSE) algorithm~\cite{sun2018learning}. Once deployed, the trained model produces high-quality decisions with reduced online complexity. This approach preserves the optimization objective while changing the computational mechanism. However, the learned mapping remains reliable only when the training data adequately represent the operating regime; shifts in network topology, traffic demand, or system constraints can cause the model to depart from the intended optimizer.

\subsection{Model-Driven Deep Learning}

The limitations of purely data-driven, black-box modules motivated the development of model-driven deep learning. Unconstrained neural networks can require large representative datasets, generalize poorly under environmental shifts, and provide limited interpretability. These limitations are especially important in wireless systems, where data collection is costly and reliable operation must remain compatible with protocol interfaces.

\begin{figure*}[!t]
    \centering
    \includegraphics[width=0.96\textwidth]{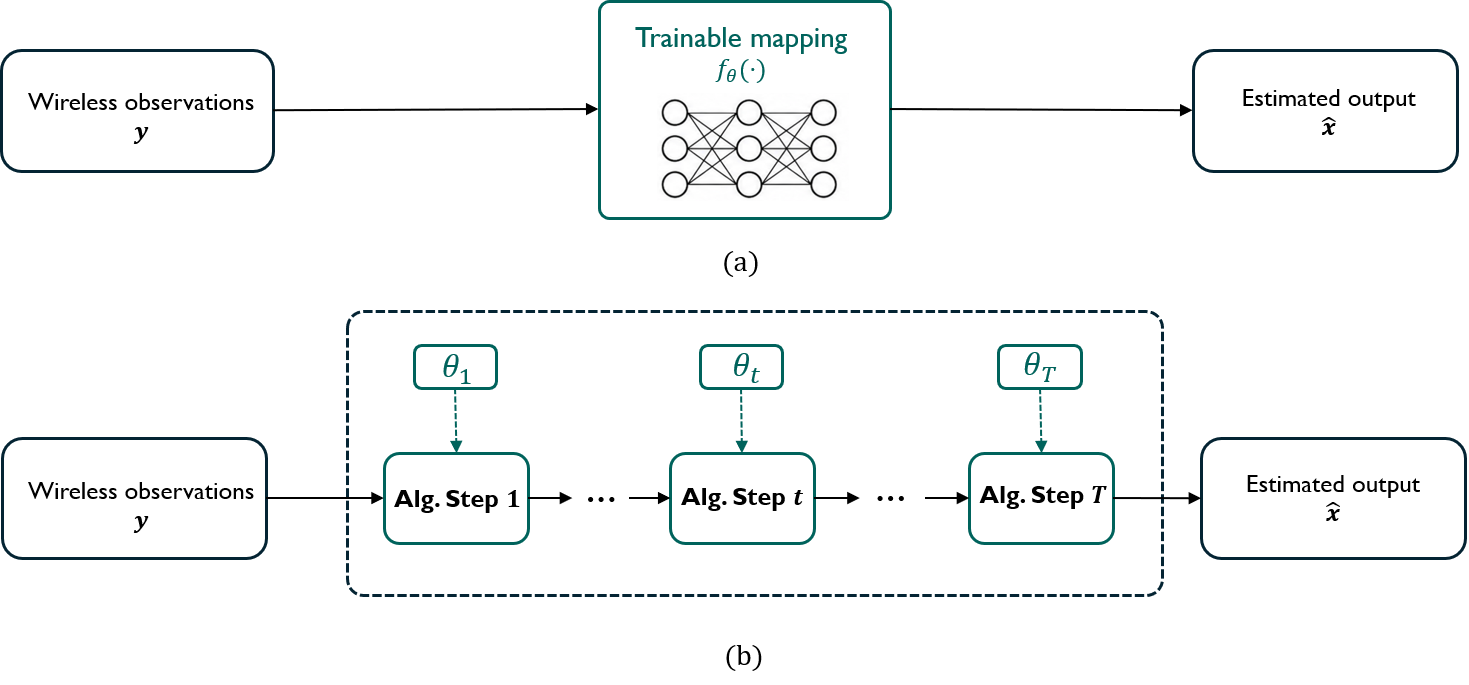}
    \caption{Two implementations of learned wireless functional modules: (a) a black-box trainable mapping from wireless observations to an estimate; and (b) model-driven deep unfolding, in which trainable parameters are embedded in successive algorithmic steps.}
    \label{fig:black_vs_unfolding}
\end{figure*}

As illustrated in Fig.~\ref{fig:black_vs_unfolding}, model-driven architectures address this tension by incorporating communication-domain knowledge into the neural topology. Rather than requiring a model to learn wireless structure from data alone, the architecture is guided by signal models, classical algorithms, and physical constraints. Learning then refines components that are difficult to specify analytically, reducing the data burden while supplying inductive biases aligned with communication physics.
Deep unfolding is a representative realization of this principle. An iterative algorithm is converted into a finite-depth neural network, where each layer corresponds to one algorithmic iteration and selected parameters are made trainable. In MIMO detection, for example, DetNet unfolds projected-gradient iterations into a trainable detector, illustrating how algorithmic structure can guide a learned architecture~\cite{samuel2019learning}.

Together, these developments show that the first role of deep learning in wireless communications was to make selected wireless functions trainable. Whether used for inference, representation, optimization, or model refinement, a learned module should provide usable information, respect the surrounding interfaces, and improve the wireless objective it is designed to serve.

\section{Redesigning and Re-normalizing Communication Goals}

The learned functional modules discussed in Section~II concern how established wireless functions are implemented. Deep learning also allows the ultimate objective of communication to be reconsidered. Rather than optimizing individual components through fixed intermediate metrics, learning can jointly optimize larger parts of the communication chain and align transmitted representations with their downstream use. Reliable transmission remains essential, but it is a means to the broader goal of making information useful to the intended recipient.

\subsection{End-to-End Communication}

End-to-end learning extends wireless design from block-level optimization to joint optimization of the transmitter--channel--receiver chain. In the autoencoder formulation, the transmitter and receiver are modeled as neural networks connected through a physical or simulated channel~\cite{oshea2017introduction}. The transmitter maps messages or source representations into channel inputs subject to physical constraints while the receiver maps channel outputs to reconstructed messages. The parameters of both networks are trained jointly by minimizing an end-to-end loss over the complete communication chain. This replaces separate intermediate objectives with a unified training objective, allowing trainable components to adapt to the final loss. Such joint adaptation can produce signaling structures and receiver mappings that are difficult to obtain under strictly modular design, particularly when the channel, hardware, or application objective deviates from idealized assumptions. The same principle underlies deep joint source--channel coding, where source representation and channel protection are learned together for end-to-end reconstruction~\cite{bourtsoulatze2019deep}.

A central technical challenge is that the channel becomes part of the training path. Standard backpropagation requires a differentiable and sufficiently accurate channel model, whereas practical wireless channels are often unknown, time varying, and affected by factors that are difficult to model explicitly. Channel surrogates, generative channel models, reinforcement learning, and over-the-air training can mitigate this limitation, but may introduce additional issues of sample efficiency, stability, and generalization~\cite{ye2020deep}. A second challenge is scalability: as block length, bandwidth, antenna dimension, or the number of users increases, end-to-end training becomes increasingly data- and computation-intensive.

\subsection{Task-Oriented and Semantic Communication}

Semantic and task-oriented communication extend the shift introduced by end-to-end learning from optimizing the communication chain to redefining the communication objective itself. Instead of treating reliable bit recovery or source reconstruction as the universal endpoint, these methods optimize transmitted representations according to downstream utility. In this sense, semantic communication can be viewed as task-driven representation learning under physical-channel constraints, where the goal is not to preserve all source details but to preserve the information that determines meaning or action. Semantics therefore takes center stage when communication is viewed in terms of the collaborative activity it supports.

\begin{figure*}[!t]
    \centering
    \includegraphics[width=0.98\textwidth]{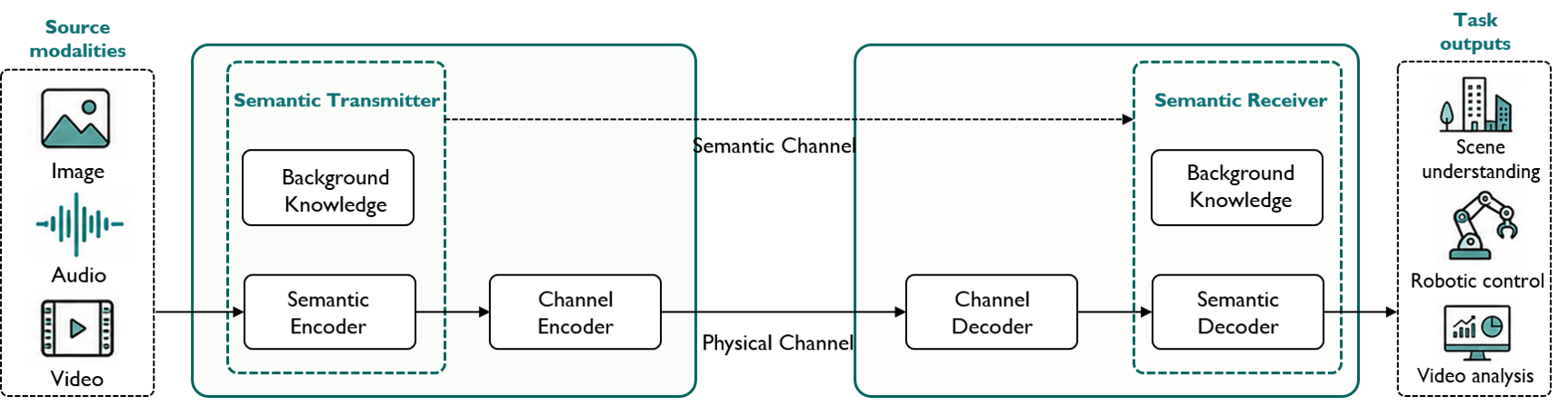}
    \caption{Representative semantic communication architecture. Source modalities are mapped to task-relevant representations through semantic and channel encoding, while shared background knowledge supports semantic decoding and downstream task outputs at the receiver.}
    \label{fig:semantic_architecture}
\end{figure*}

Figure~\ref{fig:semantic_architecture} illustrates this principle through a representative semantic communication architecture. Deep learning has enabled this direction by providing trainable representations that capture semantic structure and adapt to channel conditions. DeepSC provided an early demonstration of this idea for text transmission by optimizing semantic preservation under channel impairment rather than only bit-level accuracy~\cite{xie2021deep}. The same principle has since been extended to richer media and inference-oriented wireless systems. More recent developments incorporate external knowledge and pretrained models as shared semantic priors between transmitter and receiver, suggesting that future semantic communication may rely not only on learned compression, but also on context and reasoning to determine what information should be transmitted or inferred. Overall, semantic communication reframes wireless transmission as the selective exchange of task-relevant information, making the communication objective depend on meaning, context, and downstream use rather than signal reconstruction alone.

\subsection{Wireless Edge Intelligence}

The redefinition of communication objectives naturally extends beyond point-to-point task execution to network-scale edge intelligence. In this setting, wireless infrastructure is no longer only a transport mechanism for delivering data; it becomes a constrained fabric for distributed learning, inference, and model adaptation. The information exchanged over the air may be a model update, a feature representation, or an intermediate neural state, and its value is determined by how much it contributes to learning progress, inference quality, or efficient operation.

Federated learning provides a representative example of this shift. Edge devices compute local updates from decentralized and often statistically heterogeneous data, while the wireless network determines participation and resource use. The design objective therefore moves beyond conventional link-level metrics toward the efficiency and reliability of distributed training. Studies of federated learning over wireless networks show that local computation and wireless resource allocation should be designed jointly to reduce convergence time and energy consumption~\cite{dinh2021federated}. This establishes an important principle for edge intelligence: decisions at the physical layer and the medium access control (MAC) layer should be evaluated by their effect on model convergence and final learning performance. 
This perspective also broadens the meaning of communication efficiency. Reducing transmitted bits is useful only when the compressed information remains valuable for learning or inference. Similarly, channel-aware scheduling must be balanced with the statistical value of local updates, and this balance changes as the model evolves during training.

As edge intelligence moves toward larger models and more heterogeneous deployments, communication robustness must also be built into the learning algorithm itself. Practical devices may participate intermittently, compute with unequal resources, and exchange information over imperfect links. Methods built on exact local computation and ideal information exchange may become fragile under realistic wireless conditions. Preconditioned inexact stochastic alternating direction method of multipliers (ADMM) provides a recent example of distributed optimization for deep models under data heterogeneity and inexact local computation~\cite{zhou2026preconditioned}. This direction suggests that future wireless edge intelligence requires co-design of radio resources, model-update exchange, and learning algorithms that explicitly tolerate communication-induced inexactness.

Together, these directions show that deep learning changes not only the implementation of wireless functions, but also the objectives they serve. End-to-end learning, semantic communication, and wireless edge intelligence all move beyond fixed intermediate metrics toward communication designs shaped by downstream utility. This shift makes the value of transmitted information depend on how it supports reconstruction, inference, decision making, or learning within the broader system.

\section{Fast Adaptation, Generalization, and Efficient Wireless AI} 

The preceding sections show how deep learning can reshape wireless functional modules and communication objectives. However, these advances do not by themselves overcome the physical and operational constraints of wireless systems. Wireless environments vary across devices and deployments, while adaptation and inference must still satisfy tight signaling, computation, latency, and energy budgets. Therefore, the practical value of wireless AI depends not only on performance under a fixed training distribution, but also on whether learned models can generalize under distribution shift, adapt with limited overhead, and execute efficiently at the wireless edge.

\subsection{Fast Adaptation to Non-Stationary Environments}

Fast adaptation addresses the mismatch between offline model training and online wireless operation. A learned wireless model is typically trained over certain channel, hardware, and network conditions, whereas deployment exposes it to conditions that vary over time and across locations. The objective is therefore not only to learn a model with strong average performance, but to learn a model that can specialize rapidly when the operating condition shifts. From this perspective, different wireless environments can be viewed as related tasks that share underlying physical structure, allowing prior knowledge to be reused when only limited new pilots, measurements, or feedback samples are available.

Several learning mechanisms support this objective. Transfer learning adapts selected parts of a pretrained model, while online learning updates the model as measurements arrive. Meta-learning is particularly relevant when only a few observations are available because it learns an initialization or update rule that can be specialized rapidly. In wireless demodulation, offline and online meta-learning have enabled receiver adaptation from only a few pilots under changing end-to-end channel conditions~\cite{park2021demodulate}. More broadly, adaptation is most effective when it exploits wireless-domain structure rather than treating specialization as unconstrained retraining.

The distinctive challenge is that adaptation itself consumes wireless resources. Additional measurements, feedback, model updates, and computation reduce the net gain of learning, especially when environmental changes are temporary. A practical adaptive model must therefore improve performance after accounting for adaptation cost, remain stable under noisy observations, and avoid overreacting to short-term fluctuations. Fast adaptation is consequently not only a machine-learning technique, but a communication-constrained mechanism for maintaining learned wireless functions under non-stationary physical conditions.

\subsection{Wireless Foundation Models and Reusable Representations}

While fast adaptation reduces the cost of specialization, wireless foundation models aim to provide a reusable initialization before task-specific adaptation is required. In modern AI, foundation models are motivated by the observation that broad pretraining can improve transfer and adaptation as data and model scale increase. Applying this paradigm to wireless communications addresses a major limitation of current wireless AI: many models remain tied to a narrow task or deployment configuration. A wireless foundation model seeks to replace this fragmented workflow with a shared representation for communication, sensing, and network-intelligence tasks.

\begin{figure*}[!t]
    \centering
    \includegraphics[width=0.98\textwidth]{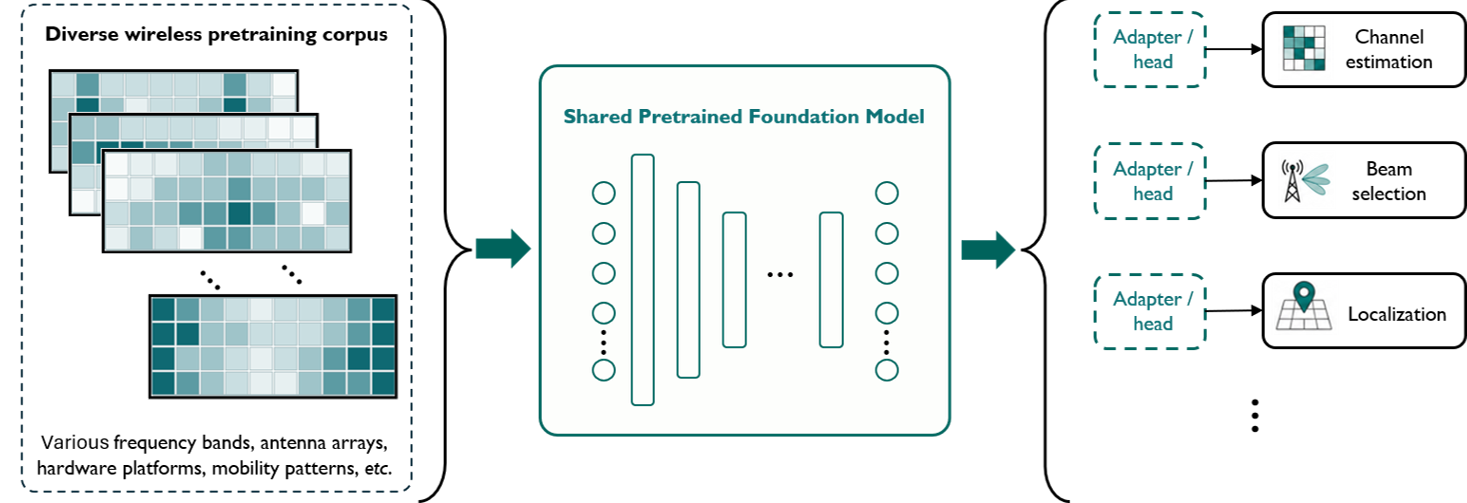}
    \caption{Reusable wireless representations through foundation-model pretraining. Diverse wireless data are used to pretrain a shared backbone, while lightweight adapters or task heads specialize the representation for downstream tasks.}
    \label{fig:foundation_model}
\end{figure*}

As illustrated in Fig.~\ref{fig:foundation_model}, this shift introduces a new wireless learning pipeline. Instead of training an independent model for each function, diverse wireless data are organized into a common pretraining corpus. A shared backbone can then be pretrained with self-supervised objectives that reconstruct masked observations and align representations across channel views. After pretraining, lightweight task heads adapt the shared representation to downstream tasks. Recent work demonstrates this approach across multiple wireless prediction tasks and previously unseen conditions~\cite{guler2026foundation}. The technical challenge is that wireless data are shaped by the measurement system itself, including frequency, antennas, hardware, mobility, deployment layout, and protocols. Transferable representations must therefore incorporate physical context and uncertainty rather than rely on scale alone, and broadly pretrained models may require compression, partitioning, or adapter-based specialization before they can satisfy edge constraints.

\subsection{Efficient Wireless AI at the Edge}

Fast adaptation and foundation models improve the flexibility of wireless AI, but they also make efficiency a central design constraint. A learned model that generalizes well may still be impractical if its inference latency, memory footprint, energy consumption, or update cost exceeds the budget of the wireless function it supports. This issue is particularly important at the edge, where devices operate under limited computation and battery resources, and where many communication decisions must be made within strict protocol time scales.

In wireless systems, the cost of inference extends beyond neural-network complexity. The relevant cost includes not only parameters, operations, and memory access, but also the pilots, feedback, signaling, and model updates required to make the learned function useful. For example, a more accurate predictor may reduce beam-training overhead or retransmissions, while a smaller model may save computation but require more frequent adaptation or produce less reliable control decisions. Efficient wireless AI is therefore the problem of matching model complexity to the time scale, resource budget, and communication role of the function being learned.

Different efficiency mechanisms address different parts of this cost. Model compression and lightweight adapters reduce size and arithmetic complexity. Adaptive-computation architectures allocate effort according to input difficulty, while split inference distributes heavier processing across devices, edge servers, and the cloud. Low-power architectures, such as spiking neural
networks, provide another direction for energy-aware inference~\cite{liu2024energy}. These techniques are especially relevant as wireless foundation models introduce larger pretrained backbones that must be specialized or compressed before they can support time-sensitive edge operation.

The central tradeoff arises at the system level rather than at the model level alone. A smaller model is not necessarily better if it increases pilot overhead, retransmissions, or control errors; a larger model may be justified if it improves robustness or reduces communication overhead. Similarly, split inference may reduce device computation while increasing signaling cost and dependence on edge connectivity. Edge-efficient wireless AI should therefore be evaluated by total system utility under computation, communication, latency, and energy budgets, rather than by compression ratio or inference accuracy alone. Together with fast adaptation and reusable representations, efficient edge inference defines the operational foundation for learned wireless systems under practical physical constraints.

\section{Challenges and Future Directions}
\label{sec:future_directions}

The preceding sections show that wireless AI is evolving from isolated learned functions toward adaptive intelligence embedded within communication systems. This evolution is now intersecting with rapid advances in foundation models, generative AI, and agentic systems, which offer new tools for representation, prediction, reasoning, and automation. The key challenge is to adapt these capabilities to wireless systems, where learned models must operate through physical channels, standardized interfaces, and strict reliability requirements. This section discusses three directions likely to shape this transition: standardization challenges for wireless AI, multimodal wireless world models, and agentic reasoning and fulfillment.

\subsection{Standardization Challenges for Wireless AI}

Standardization is essential for integrating learned models into interoperable wireless systems. Many learning-based methods are developed under implicit assumptions about their training data, operating conditions, and input-output formats. Such assumptions may be acceptable in controlled experiments, but they are insufficient when learned components interact with standardized procedures. In these settings, a learned model must be specified not only by its architecture and parameters, but also by its functional role, validity range, and behavior under uncertainty.

The central challenge is not to standardize particular neural architectures, but to define the information and procedures required for consistent integration. Future standards may need to specify model metadata, validation conditions, supported operating regimes, uncertainty indicators, and coexistence with conventional algorithms. These issues are especially important in multi-vendor and multi-device networks, where interoperability depends on common definitions of model assumptions and guarantees. Standardization should therefore provide a framework through which learned components can be evaluated, exchanged, and integrated as communication functions with explicit operational boundaries~\cite{hoydis2021toward}.

\subsection{Multimodal Wireless World Models}

Future wireless networks are expected to evolve from communication infrastructures into edge-intelligence platforms that perceive and adapt to their surrounding physical environments. Integrated sensing and communication is an important step in this direction because it allows radio signals to support both connectivity and environmental awareness. As communication measurements are combined with sensing and network context, wireless systems can move beyond isolated prediction and toward modeling the radio environment as a dynamic system shaped by physical propagation and network actions.

A multimodal wireless world model aims to provide such a predictive representation. Its purpose is not only to estimate the current channel state, but to forecast how the wireless environment may evolve and respond to candidate control decisions. This capability could support proactive communication control and simulation-based evaluation before actions are applied to the live network. To be useful, the model must remain physically grounded: its predictions and generated scenarios should respect propagation behavior, uncertainty, causality, and operational constraints. Wireless foundation models provide an initial step toward reusable representations of the radio environment, but a full world model must additionally capture action-conditioned dynamics and uncertainty~\cite{guler2026foundation}.

\subsection{Agentic Reasoning and Fulfillment}

Agentic reasoning and fulfillment represent a step from predictive wireless AI toward closed-loop wireless intelligence. Rather than executing a fixed mapping from observations to actions, an agentic system follows a perceive--reason--act cycle: it observes network and environmental context, reasons over service objectives, and acts through trusted communication-control tools. This framework introduces memory, knowledge, tool use, and planning into wireless decision making, enabling agents to interpret intent, diagnose conditions, coordinate network functions, and interact with other agents or human operators. Recent work has framed this progression as a transition from task-specific adaptation toward increasingly autonomous wireless intelligence~\cite{liang2026llm}. In this broader vision, fulfillment extends the loop across the communication process, from message origination to contact establishment through intelligent switching and ultimately to the delivery of semantically critical messages. Intelligent switching allows agents, edge intelligence, and the network to identify the intended receiver and establish the connection without requiring the sender to specify the destination through a conventional numbering plan.

\begin{figure*}[!t]
    \centering
    \includegraphics[width=0.88\textwidth]{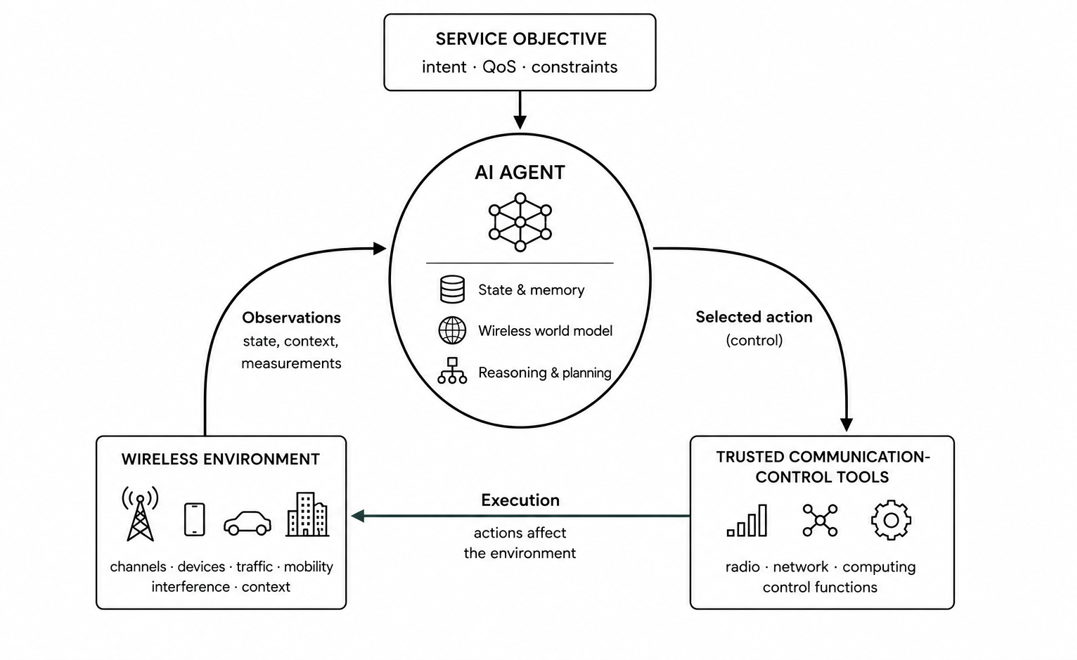}
    \caption{Agentic wireless intelligence as a closed loop among the service objective, an AI agent, trusted communication-control tools, and the wireless environment. The agent reasons from observations, selects actions for execution, and receives feedback through the resulting changes in the environment.}
    \label{fig:agentic_loop}
\end{figure*}

Figure~\ref{fig:agentic_loop} summarizes this closed-loop view. The central challenge is to make this flexibility compatible with the predictability required by wireless infrastructure. Large language models may provide reasoning and coordination capabilities, but radio actions affect interference, reliability, and other control loops; therefore, agentic communication must operate within explicit boundaries. Standardized interfaces can define what an agent is allowed to observe or modify, while wireless world models can provide predictive grounding before actions are applied to the live network. Future research should emphasize verifiable tool use, uncertainty awareness, interpretable decision traces, and fallback mechanisms, so that wireless agents can adapt flexibly while remaining physically grounded and interoperable.

\section{Conclusion}

The first decade of deep learning for wireless communications has shown that learning can influence wireless-system design at multiple levels. It can learn functional modules, redesign and re-normalize communication goals, and support more generalizable operation through adaptation, reusable representations, and efficient inference. Looking ahead, the central challenge is to turn these capabilities into dependable elements of wireless infrastructure that support communication anytime and anywhere, through any appropriate means. This requires learned models whose roles, interfaces, and operating assumptions are clearly specified, whose behavior remains reliable under relevant wireless conditions, and whose implementation respects the latency, energy, and protocol constraints of practical networks.

\bibliographystyle{IEEEtran}
\bibliography{references}

\end{document}